      [1999/12/01 v1.4c Il Nuovo Cimento]
\documentclass{cimento}

\newcommand{\bad}{\begin{array}{ccc}}
\newcommand{\baz}{\begin{array}{cc}}
\newcommand{\ba}{\begin{array}{c}}
\newcommand{\ea}{\end{array}}
\newcommand{\nn}{\nonumber}

\title{Mass hierarchy and flavour mixing from discrete symmetries}
\author{Yin Lin\from{ins:a}}
\instlist{\inst{ins:a} Universit\`a di Padova, INFN Sezione di Padova}
\PACSes{\PACSit{14.60.Pq}{Neutrino mass and mixing}\PACSit{11.30.Hv}{Flavour symmetries}}
\begin{document}

\maketitle

\begin{abstract}
We consider a class of discrete flavour symmetries for leptons based on the group $S_3$ and $A_4$
with an hybrid breaking pattern. The aim is to construct models in which the 
same flavon fields producing the mixing pattern are also responsible for the mass hierarchy.
\end{abstract}

\section{Introduction}
Nowadays continuous improvement on the knowledge of neutrino oscillation parameters  
makes desirable a neutrino model building going beyond the mere fitting procedure.
The present data \cite{data}, at 1$\sigma$:
\begin{equation}
\theta_{12}=(34.5\pm 1.4)^o~~~,~~~~~\theta_{23}=(42.3^{+5.1}_{-3.3})^o~~~,~~~~~\theta_{13}=(0.0^{+7.9}_{-0.0})^o~~~,
\label{angles}
\end{equation}
are fully compatible with the so-called Tri-Bimaximal (TB) mixing pattern
which corresponds to a maximal  $\theta_{23}$, an zero  $\theta_{13}$ and
a ``magic'' value of solar angle:  $\sin^2\theta_{12}=1/3$.

It is well known that the lepton mixing angles can be understood by a mechanism of 
vacuum misalignment in flavour space
occurring in theories with non-abelian flavour symmetries \cite{review}.
Also the charged lepton mass hierarchy can be achieved 
via spontaneous breaking of the flavour symmetry.
However, in most cases, a separate component of the flavour group is exploited to this purpose.
Quite frequently the flavour group is of the type $D\times U(1)_{\rm{FN}}$ where
$D$ is a discrete component that controls the mixing angles and $U(1)_{\rm{FN}}$  \cite{FN}
is an abelian continuous symmetry that describes the mass hierarchy.

In the present talk, we will consider a class of flavour symmetries in which the same flavon fields
producing the mixing pattern are also responsible for the mass hierarchy
employing an hybrid breaking pattern \cite{works}.
The idea is to selectively couple charged leptons and neutrinos to two different sets of flavons, 
$\Phi_e$ and $\Phi_\nu$, respectively. The VEV of $\Phi_\nu$ breaks $G$ down to a
residual symmetry in the neutrino sector preferably containing the $\nu_\mu-\nu_\tau$ 
exchange symmetry as indicated by the oscillation data.
While, the VEV of $\Phi_e$ would break $G$ down to a different subgroup, 
maximally breaking the previous $\nu_\mu-\nu_\tau$ symmetry, guaranteeing a hierarchical
and quasi diagonal matrix $m_l$.

\section{Neutrino $\nu_\mu -\nu_\tau$ symmetry, charged lepton hierarchy and $S_3$}

In this section we will focus on the approximately vanishing 
values of $\theta_{13}$ and $\theta_{23}- \pi/4$.
Given an arbitrary choice of charged lepton mass matrix $m_l$ and effective neutrino mass matrix 
$m_{\nu}$, any change of basis in the generation space modifies the form of 
$m_l$ and $m_{\nu}$, but does not change the physics.
We can exploit this freedom to render diagonal the charged lepton mass matrix $m_l$:
\begin{equation} 
\label{flavourbasis}
m'_l= {\rm {diag}} (m_e, m_{\mu}, m_{\tau}), \qquad m'_{\nu}=U_{PMNS}^* {\rm{diag}} (m_1, m_2, m_3)U^\dagger_{PMNS}~~.
\end{equation}
In this basis, called flavour basis, the effective neutrino mass matrix is completely determined by the measurable quantities $m_i$ and $U_{\rm{PMNS}}$. 
In the limit where both $\theta_{13}$ and $\theta_{23}- \pi/4$ vanish, it is easy to verify that
$m_\nu'$ exhibits an exact $\nu_\mu-\nu_\tau$ exchange symmetry. 
However, since $\nu_\mu$ and $\nu_\tau$ 
are members of the electroweak doublets, 
a naive extension of such a symmetry to include the charged leptons
$\mu$, $\tau$ might be in contrast with the large mass hierarchy $m_\mu\ll m_\tau$.
This problem is completely solved in the first paper of \cite{works} by
using an hybrid symmetry pattern based
on the flavour symmetry $S_3$. 

$S_3$ is group of permutations of three distinct objects and is the smallest non-abelian symmetry group.
The six elements of the $S_3$ group can be generated by $S$ and $T$ with the 
following unitary representations (Rep): 
\begin{equation} \label{gen}
\left\{ \bad 1  & S=1 & T=1 \\ 1' & S=-1 & T=1\\ 
\ea \right.
\qquad
2~~~S = \left( \baz 
0 & 1 \\[0.2cm] 
1 & 0 \\ 
               \ea   \right)~                
T = \left( \baz 
\omega & 0 \\[0.2cm] 
0 & \omega^2 \\ 
               \ea   \right) ~.
\end{equation}
$S$ in the two-dimensional Rep corresponds to an interchange symmetry of the two components 
of a $S_3$ doublet. The tensor products involving pseudo-singlets are given 
by $1'  \times 1' =1 $ and $1' \times 2=2$. While the product of two doublets is given by 
$2\times 2=2+1+1'$. Given two doublets $\psi=(\psi_1,\psi_2)$ 
and $\varphi=(\varphi_1,\varphi_2)$, it is easy to see that
\begin{equation} 
\label{tensorprod1s3}
\ba
\psi_1\varphi_2+\psi_2\varphi_1 \in 1 \\
\psi_1\varphi_2-\psi_2\varphi_1 \in 1' \\
\ea
\qquad
 \left( 
 \ba
 \psi_2\varphi_2 \\
  \psi_1\varphi_1 \\
  \ea
  \right)
  \in 2
 \end{equation}

The previous notion on the $S_3$ group is sufficient to construct a simple 
 model for leptons.
The left-handed doublets transform as $(1+2)$ of $S_3$ and we will call 
$l_e=(\nu_e,e)$ the invariant singlet and $D_l=(l_\mu, l_\tau)$ the $S_3$ doublet. 
The right-handed charged leptons $e^c, \mu^c, \tau^c$
are all in the non-trivial singlet representation $1'$. The $S_3$ flavour symmetry is spontaneously
broken at a scale $\sim \rm{VEV} \ll \Lambda$, $\Lambda$ being the cutoff scale,
by two doublets $\varphi_e, \varphi_\nu$ and a singlet $\xi$ 
which are all gauge singlets.
The flavon fields will develop VEVs of the type
\begin{equation}
\langle \varphi_e \rangle \propto (1,0)~,  \qquad \langle \varphi_\nu \rangle= (1,1)~,
\qquad \langle \xi \rangle \ne 0~~~.
\label{vevsnu}
\end{equation} 
It is possible to impose an extra abelian symmetry
$Z_3$ in such a way that  $\varphi_\nu$ and $\xi$ couple only to the neutrino sector and
$\varphi_e$ to the charged lepton sector. 
Since $\langle \varphi_\nu \rangle$ is preserved by $S$, 
we immediately conclude that in the neutrino sector $S_3$ is broken down to a $Z_2$ subgroup. 
Being $D_l$ doublet of $S_3$,
this residual $Z_2$ parity will exactly lead to the $\nu_\mu-\nu_\tau$ exchange symmetry.
Concerning the charged lepton sector, the VEV of $\varphi_e$ breaks the parity symmetry 
generated by $S$ in a maximal way, since 
\begin{equation}
\langle \varphi_e \rangle^\dagger S \langle \varphi_e \rangle=0~~~.
\end{equation}
The singlets $e^c$, $\mu^c$ and $\tau^c$ should
couple to $(D_l\varphi_e)'$,  $(D_l\varphi_e\varphi_e)'$
and $l_1(\varphi_e\varphi_e\varphi_e)'$ at the leading orders. 
From the $S_3$ multiplication rules given in
 (\ref{tensorprod1s3}) we see that the last combinations select respectively
 $e$, $\mu$ and $\tau$ after the electroweak and flavour symmetry breaking.
As a consequence $m_\tau$, $m_\mu$, $m_e$ get their first non-vanishing contribution
at the order $\langle \varphi_e\rangle/\Lambda$, $(\langle \varphi_e\rangle/\Lambda)^2$, $(\langle \varphi_e\rangle/\Lambda)^3$, respectively. 
Then we obtain the correct charged lepton hierarchy 
assuming $\langle \varphi_T\rangle / \Lambda \sim \lambda_c^2 
$, being $\lambda_c$ the Cabibbo angle. 

\section{TB mixing and charged lepton hierarchy from $A_4$}

In this section we extend the previous construction to describe the 
TB mixing pattern by an hybrid symmetry breaking of $A_4$ \cite{works},
improving some aspects of the original proposal of \cite{TBA4}.
The generators of $A_4$, $S$ and $T$, have the following Reps:
\begin{equation}
\begin{array}{lll}
1&S=1&T=1\\
1'&S=1&T=\omega^2\\
1''&S=1&T=\omega
\end{array} 
\qquad
3~~~
T=\left(
\begin{array}{ccc}
1&0&0\\
0&\omega^2&0\\
0&0&\omega
\end{array}
\right)~
S=\frac{1}{3}
\left(
\begin{array}{ccc}
-1&2&2\cr
2&-1&2\cr
2&2&-1
\end{array}
\right) \nn
\end{equation}
From (\ref{flavourbasis}), one can generally show that the most general mass matrix 
in the flavour basis leading to TB mixing
obeys the ``magic'' symmetry  $G_S \simeq Z_2$ generated by $S$ in addition 
to the $\nu_\mu-\nu_\tau$ exchange symmetry analyzed in the previous section.

We assign the lepton doublets $l_i$ ($i=e, \mu, \tau$) to the triplet $A_4$ 
representation and the lepton singlets $e^c, \mu^c, \tau^c \sim 1$.
The symmetry breaking sector consists of the scalar fields $(\varphi_T,\varphi_S,\xi)$,
transforming as $(3,3,1)$ of $A_4$. Under certain conditions, the minimization 
of the scalar potential naturally leads to the following VEV alignment:
\begin{equation}
\langle \varphi_T\rangle\propto (0, 1,0)~,\qquad \langle \varphi_S\rangle\propto (1,1,1)~,
\qquad \langle \xi\rangle\ne 0~~~.
\label{va1}
\end{equation}
In the neutrino sector $A_4$ is broken by $(\varphi_S, \xi)$ down to $G_S$.
The absence of the scalar singlets $1'$ and $1''$ in the neutrino sector implies that,
in addition to $G_S$, the resultant neutrino mass matrix is also automatically
symmetric under the exchange of the second and third generations. 
Then the residual symmetry
in the neutrino sector is enhanced and
imposes $U_{\rm{PMNS}}$ to be of the TB form independently
from the mass eigenvalues.
In the charged lepton sector, the VEV of $\varphi_T$ breaks the $\nu_\mu-\nu_\tau$ 
exchange symmetry in a maximal way similar to the case of
$S_3$. The masses $m_e$, $m_\mu$ and $m_\tau$
arise at the order $\langle \varphi_T\rangle /\Lambda$, $(\langle \varphi_T\rangle /\Lambda)^2$
and $(\langle \varphi_T\rangle /\Lambda)^3$ respectively leading to a hierarchical
mass spectrum.

\acknowledgments
I wish to thank the organizers of IFAE 2008 and F. Feruglio for collaborations.

\end{document}